\documentclass[11pt]{article}
\usepackage{amsmath,amssymb}
\usepackage{mathtools} 
\usepackage{array}
\newcount\figureno     \figureno=0
\newdimen\figdim       \figdim=70mm
\def\figureinc{%
   \global\advance\figureno by 1%
}
\def\figcaption#1#2#3{\hbox to #2{\hss{\vbox{\hsize=#2 \parindent=0pt
        {\bf Figure \number\figureno#3 :\ }#1}}\hss}
}

\begin{document}
\baselineskip 100pt
\renewcommand{\baselinestretch}{1.5}
\renewcommand{\arraystretch}{0.666666666}
{\large
\parskip.2in
\numberwithin{equation}{section}

\newcommand{\opa}{\overline{\partial}}   
\newtheorem{theorem}{Theorem}[section]
\newtheorem{lemme}{Lemma}[section]
\newtheorem{proposition}{Proposition}[section]
\newtheorem{corollaire}{Corollary}[section]
\newtheorem{exemple}{Example}[section]

\newcommand{\be}{\begin{equation}}
\newcommand{\ee}{\end{equation}}
\newcommand{\ben}{\begin{equation*}}
\newcommand{\een}{\end{equation*}}
\newcommand{\eqalinb}{\begin{eqnarray}}
\newcommand{\eqaline}{\end{eqnarray}}
\newcommand{\br}{\bar}
\newcommand{\fr}{\frac}
\newcommand{\lm}{\lambda}
\newcommand{\ra}{\rightarrow}
\newcommand{\al}{\alpha}
\newcommand{\bt}{\beta}
\newcommand{\z}{\zeta}
\newcommand{\pa}{\partial}
\newcommand{\hs}{\hspace{5mm}}
\newcommand{\up}{\upsilon}
\newcommand{\bigb}{\hspace{7mm}}
\newcommand{\dg}{\dagger}
\newcommand{\vphi}{\vec{\varphi}}
\newcommand{\ve}{\varepsilon}
\newcommand{\acc}{\\[3mm]}
\newcommand{\dl}{\delta}
\newcommand{\sdil}{\ensuremath{\rlap{\raisebox{.15ex}{$\mskip
6.5mu\scriptstyle+ $}}\subset}}
\newcommand{\sdir}{\ensuremath{\rlap{\raisebox{.15ex}{$\mskip
6.5mu\scriptstyle+ $}}\supset}}
\def\tablecap#1{\vskip 3mm \centerline{#1}\vskip 5mm}
\def\p#1{\partial_#1}
\newcommand{\pd}[2]{\frac{\partial #1}{\partial #2}}
\newcommand{\pdn}[3]{\frac{\partial #1^{#3}}{\partial #2^{#3}}}
\def\DP#1#2{D_{#1}\varphi^{#2}}
\def\dP#1#2{\partial_{#1}\varphi^{#2}}
\def\xh{\hat x}
\newcommand{\Ref}[1]{(\ref{#1})}
\def\ld{\,\ldots\,}

\def\C{{\mathbb C}}
\def\Z{{\mathbb Z}}
\def\R{{\mathbb R}}
\def\mod#1{ \vert #1 \vert }
\def\chapter#1{\hbox{Introduction.}}
\def\Sin{\hbox{sin}}
\def\Cos{\hbox{cos}}
\def\Exp{\hbox{exp}}
\def\Ln{\hbox{ln}}
\def\Tan{\hbox{tan}}
\def\Cot{\hbox{cot}}
\def\Sinh{\hbox{sinh}}
\def\Cosh{\hbox{cosh}}
\def\Tanh{\hbox{tanh}}
\def\Asin{\hbox{asin}}
\def\Acos{\hbox{acos}}
\def\Atan{\hbox{atan}}
\def\Asinh{\hbox{asinh}}
\def\Acosh{\hbox{acosh}}
\def\Atanh{\hbox{atanh}}
\def\frac#1#2{{\textstyle{#1\over #2}}}

\def\ph{\varphi_{m,n}}
\def\phl{\varphi_{m-1,n}}
\def\phr{\varphi_{m+1,n}}
\def\varphil{\varphi_{m-1,n}}
\def\varphir{\varphi_{m+1,n}}
\def\varphit{\varphi_{m,n+1}}
\def\varphib{\varphi_{m,n-1}}
\def\pht{\varphi_{m,n+1}}
\def\phb{\varphi_{m,n-1}}
\def\phbl{\varphi_{m-1,n-1}}
\def\phbr{\varphi_{m+1,n-1}}
\def\phtl{\varphi_{m-1,n+1}}
\def\phtr{\varphi_{m+1,n+1}}
\def\u{u_{m,n}}
\def\ul{u_{m-1,n}}
\def\ur{u_{m+1,n}}
\def\ut{u_{m,n+1}}
\def\ub{u_{m,n-1}}
\def\utr{u_{m+1,n+1}}
\def\ubl{u_{m-1,n-1}}
\def\utl{u_{m-1,n+1}}
\def\ubr{u_{m+1,n-1}}
\def\v{v_{m,n}}
\def\vl{v_{m-1,n}}
\def\vr{v_{m+1,n}}
\def\vt{v_{m,n+1}}
\def\vb{v_{m,n-1}}
\def\vtr{v_{m+1,n+1}}
\def\vbl{v_{m-1,n-1}}
\def\vtl{v_{m-1,n+1}}
\def\vbr{v_{m+1,n-1}}

\def\U{U_{m,n}}
\def\Ul{U_{m-1,n}}
\def\Ur{U_{m+1,n}}
\def\Ut{U_{m,n+1}}
\def\Ub{U_{m,n-1}}
\def\Utr{U_{m+1,n+1}}
\def\Ubl{U_{m-1,n-1}}
\def\Utl{U_{m-1,n+1}}
\def\Ubr{U_{m+1,n-1}}
\def\V{V_{m,n}}
\def\Vl{V_{m-1,n}}
\def\Vr{V_{m+1,n}}
\def\Vt{V_{m,n+1}}
\def\Vb{V_{m,n-1}}
\def\Vtr{V_{m+1,n+1}}
\def\Vbl{V_{m-1,n-1}}
\def\Vtl{V_{m-1,n+1}}
\def\Vbr{V_{m+1,n-1}}
\def\tr{{\rm tr}\,}

\def\a{\alpha}
\def\b{\beta}
\def\g{\gamma}
\def\d{\delta}
\def\ep{\epsilon}
\def\e{\varepsilon}
\def\z{\zeta}
\def\t{\theta}
\def\k{\kappa}
\def\l{\lambda}
\def\s{\sigma}
\def\f{\varphi}
\def\w{\omega}
\def\v{{\hbox{v}}}
\def\u{{\hbox{u}}}
\def\x{{\hbox{x}}}

\newcommand{\ie}{{\it i.e.}}
\newcommand{\cmod}[1]{ \vert #1 \vert ^2 }
\newcommand{\cmodn}[2]{ \vert #1 \vert ^{#2} }
\newcommand{\nhat}{\mbox{\boldmath$\hat n$}}
\nopagebreak[3]
\bigskip

\title{ \bf $\mathbb{C}P^{2S}$ sigma models described through\\ hypergeometric orthogonal polynomials}
\vskip 1cm

\bigskip
\author{
N. Crampe\thanks{email
address: crampe1977@gmail.com}
\\
Centre de Recherches Math{\'e}matiques, Universit{\'e} de Montr{\'e}al,\\
C. P. 6128, Succ.\ Centre-ville, Montr{\'e}al, (QC) H3C 3J7,
Canada\\
Institut Denis Poisson, Universit{\'e} de Tours - Universit{\'e} d'Orl{\'e}ans, \\Parc de Grandmont, 37200, Tours, France
\acc 
A.~M. Grundland\thanks{email address: grundlan@crm.umontreal.ca}
\\
Centre de Recherches Math{\'e}matiques, Universit{\'e} de Montr{\'e}al,\\
C. P. 6128, Succ.\ Centre-ville, Montr{\'e}al, (QC) H3C 3J7,
Canada\\ Universit\'{e} du Qu\'{e}bec, Trois-Rivi\`{e}res, CP500 (QC) G9A 5H7, Canada 
} \date{}

\maketitle
\begin{abstract}
The main objective of this paper is to establish a new connection between the Hermitian rank-1 projector solutions of the Euclidean $\mathbb{C}P^{2S}$ sigma model in two dimensions 
and the particular hypergeometric orthogonal polynomials called Krawtchouk polynomials. We show that any Veronese subsequent analytical solutions of the $\mathbb{C}P^{2S}$ model, defined on the 
Riemann sphere and having a finite action, can be explicitly parametrised in terms of these polynomials. 
We apply these results to the analysis of surfaces associated with $\mathbb{C}P^{2S}$ models defined using the generalised Weierstrass formula for immersion. We show that these surfaces
are homeomorphic to spheres in the $\mathfrak{su}(2s+1)$ algebra, and express several other geometrical characteristics in terms of the Krawtchouk polynomials.
Finally, a connection between the $\mathfrak{su}(2)$ spin-s representation and the $\mathbb{C}P^{2S}$ model is explored in detail. 
\end{abstract}
\newpage



\vspace{3mm}


\section{Introduction}
\noindent Among the various sigma models, the one which has been the most studied is the completely integrable two-dimensional Euclidean $\mathbb{C}P^{2S}$ sigma model 
defined on the extended complex plane $\mathbb{S}^2$ having finite action. 
This subject was first analysed in the work of Din and Zakrzewski \cite{Din1,Din2}, next by Borchers and Garber \cite{Borchers}, Sasaki \cite{Sasaki} 
and later discussed by Eells and Wood \cite{Eells}, Uhlenbeck \cite{Uhlembeck}. 
It was shown by Mikhailov and Zakharov \cite{Mikhailov,Zakharov} that the Euler--Lagrange (EL) equations can be reformulated as a linear spectral problem which proves
to be very useful for the construction and analysis of explicit multi-soliton solutions of the $\mathbb{C}P^{2S}$ model. 
The main feature of this model is that all rank-1 Hermitian projector solutions of this model are obtained through successive applications of a creation operator.
This rich yet restrictive character makes the $\mathbb{C}P^{2S}$ sigma models a rather special and interesting object to study. A number of attempts to generalise the 
$\mathbb{C}P^{2S}$ models and their various applications can be found in the recent literature of the subject 
(see e.g. \cite{Bobenko,Chern,Guest,Helein,Manton,Babelon,Polyakov,Grun17} and references therein).

\noindent In this paper, we show that a special class of rank-1 Hermitian projector solutions of the $\mathbb{C}P^{2S}$ sigma models are represented in 
terms of the Krawtchouk orthogonal polynomials \cite{kra}. We also find new explicit analytical expressions for the sequence (called the Veronese sequence) of solutions
of the $\mathbb{C}P^{2S}$ sigma model. The explicit parametrisation of solutions of this model in terms of the Krawtchouk polynomials has not been previously found. 
These solutions may in turn be used to study the immersion functions of two-dimensional (2D)-soliton surfaces. This task has been accomplished by introducing a geometric setting for 
a given set of $2s+1$ rank-1 projector solutions of the $\mathbb{C}P^{2S}$ sigma models written in terms of conservation laws. The latter enable us to construct the so-called 
generalised Weierstrass formula for the immersion of the 2D-surfaces in the $\mathfrak{su}(2s+1)$ algebra. Consequently, the analytical results obtained in this paper allow us to 
explore some geometrical properties of these surfaces, including the Gaussian and the mean curvatures and some global characteristics such as the Willmore functional, the topological charge 
and the Euler--Poincar\'e characters. We show that for any Veronese subsequent solutions the topological charge of the 2D-surfaces are integers, 
while their Euler--Poincar\'e characters remain constant and equal to 2. It is shown that all Gaussian curvatures are positive for all 2D-surfaces associated with the $\mathbb{C}P^{2S}$ sigma models and therefore
these surfaces are homeomorphic to spheres immersed in the Euclidean space $\mathbb{R}^{4s(s+1)}$.

\noindent This paper is organized as follows. Section 2 contains a brief account of the projector formalism associated with the $\mathbb{C}P^{2S}$ sigma model.  
Section 3 is devoted to the construction and investigation of Veronese sequence solutions of the $\mathbb{C}P^{2S}$ model which are expressed in terms of the Krawtchouk orthogonal polynomials. 
These results are then used in Section 4 to construct the explicit form of the Clebsh-Gordan coefficients associated with this model. 
In Section 5, we discuss in detail the $\mathfrak{su}(2)$ spin-$s$ representation associated with the Veronese sequence solutions of the $\mathbb{C}P^{2S}$ model. We construct new algebraic recurrence relations from a given holomorphic solution of the model, which are simpler than the known analytic relations.
In Section 6, we present a geometric formulation for Veronese immersions of 2D-surfaces associated with this model.  Section 7 contains possible suggestions concerning further developments

\section{Projector formalism}
\subsection{Preliminaries on the $\mathbb{C}P^{2S}$ sigma model}
\noindent In the study of $\mathbb{C}P^{2S}$ models on Euclidean space, we search for the maps 
\begin{equation}
\mathbb{S}^2\supset\Omega \ni \xi_\pm = \xi^1\pm i\xi^2 \mapsto z = (z_0, z_1, ..., z_{2s})\in \mathbb{C}^{2s+1}\backslash \{\emptyset\}, 
\end{equation}
where $z^\dagger z = 1$, which are stationary points of the action functional \cite{Zakrzewski}
\begin{equation}\label{eq:action_func}
\mathcal{A}(z) = \sum_{\mu = 1}^2 \iint_\Omega \left(D_\mu z\right)^\dagger D_\mu z\; d\xi^1 d\xi^2,
\end{equation} 
defined on a simply connected open subset $\Omega$ of the extended complex plane $\mathbb{S}^2=\mathbb{C}\cup\{\infty\}$.
The corresponding EL equations are given by
\begin{equation}\label{eq:ELequations}
\sum_{\mu = 1}^2 D_\mu D_\mu z + z\left(\left(D_\mu z\right)^\dagger D_\mu z\right)=0,
\end{equation}
where $D_\mu$ are the covariant derivatives defined by
\begin{equation}
D_\mu z = \partial_\mu z - \left( z^\dagger \partial_\mu z   \right)z, \qquad \partial_\mu = \frac{\pa}{\pa \xi^\mu},\quad \mu=1,2.
\end{equation}
The action functional (\ref{eq:action_func}) is invariant under the local $\mathcal{U}(1)$ transformation induced by $k: \Omega \rightarrow \mathbb{C}\backslash\{\emptyset\}$ with $|k| = 1$, since $D_\mu (kz) = kD_\mu(z)$ holds. 
In what follows, it is convenient to use the homogeneous variables
\begin{align}\label{eq:homog_var}
\mathbb{S}^2\supset\Omega \ni \xi_\pm &= \xi^1\pm i\xi^2 \mapsto f \in \mathbb{C}^{2s+1}\backslash \{\emptyset\}, 
\end{align}
and expand the model so that the action (\ref{eq:action_func}) becomes \cite{Zakrzewski}
\begin{equation}\label{eq:action_int2}
\mathcal{A}(f) = \sum_{\mu = 1}^2 \iint_\Omega \frac{\displaystyle \left(D_\mu f\right)^\dagger D_\mu f}{\displaystyle f^\dagger f}\; d\xi^1 d\xi^2, 
\end{equation}
where
\begin{equation}
D_\mu f = \pa_\mu f - \frac{\displaystyle f^\dagger \pa_\mu f}{\displaystyle f^\dagger f}f.
\end{equation}
The action integrals (\ref{eq:action_func}) and (\ref{eq:ELequations}) are consistent with the relation
\begin{equation}
z = \frac{\displaystyle f}{\displaystyle |f|}, \qquad |f|=\left(f^\dagger f\right)^{1/2},
\end{equation}
which links the inhomogeneous coordinates $z$ with the functions $f$ as homogeneous coordinates of the model. Note that for any functions 
$f,g: \Omega \rightarrow \mathbb{C}^{2s+1}\backslash \{\emptyset\}$ such that $f = kg$ for some $k: \Omega \rightarrow \mathbb{C}\backslash \{\emptyset\} $, the action functions remain the same, i.e. $\mathcal{A}(f) = \mathcal{A}(kg) = \mathcal{A}(g)$. Using the standard notation for the complex derivatives $\pa$ and $\overline{\pa}$ with respect to $\xi_+$ and $\xi_-$, i.e.
\begin{equation}
\pa = \frac{1}{2}\left(\frac{\pa}{\pa \xi^1}-i\frac{\pa}{\pa \xi^2}\right), \qquad \overline{\pa} =  \frac{1}{2}\left(\frac{\pa}{\pa \xi^1}+i\frac{\pa}{\pa \xi^2}\right),
\end{equation}
we obtain that the $\mathbb{C}P^{2S}$ model equations expressed in terms of the $f$'s satisfy an unconstrained form of the EL equations
\begin{equation}\label{eq:ELeq2}
\left(\mathbf{1}_{2s+1}-\frac{f \otimes f^\dagger}{f^\dagger f}\right)\left[\pa\opa f - \frac{1}{f^\dagger f}\left(\left(f^\dagger \opa f\right)\pa f +\left(f^\dagger \pa f\right)\opa f \right)\right] = 0,
\end{equation}
where $\mathbf{1}_{2s+1}$ is the $(2s+1)\times (2s+1)$ identity matrix. 
In this paper we consider a special class of holomorphic solutions of \eqref{eq:ELeq2} called the Veronese sequence \cite{Bolton}
\begin{equation}\label{eq:Veronese}
f_0 = \left(1, \sqrt{\binom{2s}{1}}\xi_+, ..., \sqrt{\binom{2s}{r}}\xi_+^r, ..., \xi_+^{2s}\right)\in \mathbb{C}^{2s+1}\backslash \{\emptyset\}.
\end{equation}
It was shown \cite{Din1,Din2} that for finite action integrals (\ref{eq:action_int2}), subsequent solutions $f_k$ can be obtained by acting with the creation (annihilation) operators. In terms of the homogeneous coordinates $f_k$, new solutions of the EL equations (\ref{eq:ELeq2}) are obtained by the recurrence relations
\begin{align}\nonumber
f_{k+1} &= P_+(f_k) = \left(\mathbf{1}_{2s+1} - \frac{f_k\otimes f_k^\dagger}{f_k^\dagger f_k}\right)\pa f_k,\\\label{eq:recurrence}
f_{k-1} &= P_-(f_k) = \left(\mathbf{1}_{2s+1} - \frac{f_k\otimes f_k^\dagger}{f_k^\dagger f_k}\right)\opa f_k,\qquad 0\leq k \leq 2s,
\end{align}
where $P_\pm$ are raising and lowering operators with the properties
\begin{equation}
P_\pm^0 = \mathbf{1}_{2s+1}, \qquad P_\pm^{2s+1}(f_k) = 0, \qquad 0\leq k \leq 2s.
\end{equation}
This procedure allows us to construct three classes of solutions; namely holomorphic $f_0$, antiholomorphic $f_{2s}$ and mixed solutions $f_k, \; 1\leq k \leq 2s-1$. Let us remark that the homogeneous coordinates $\{f_k\}$ are orthogonal, i.e. $f_k^{\dagger}f_l=0$ if $k\neq l$.
\subsection{Rank-1 Hermitian projectors}
\noindent The most fruitful approach to the study of the general properties of the $\mathbb{C}P^{2S}$ model has been formulated through descriptions of the model in terms of rank-1 Hermitian projectors \cite{Goldstein1}. A matrix $P_k(\xi_+, \xi_-)$ is said to be a rank-1 Hermitian projector if
\begin{equation}\label{eq:proj_constraints}
P_k^2 = P_k, \qquad P_k^\dagger = P_k, \qquad tr(P_k) = 1.
\end{equation}
The target space of the projector $P_k$ is determined by a complex line in $\mathbb{C}^{2s+1}$, i.e. by a one-dimensional vector function $f_k(\xi_+, \xi_-)$, given by
\begin{equation}\label{eq:projector}
P_k = \frac{\displaystyle f_k\otimes f_k^\dagger}{\displaystyle f_k^\dagger f_k},
\end{equation}
where $f_k$ is the mapping given by (\ref{eq:homog_var}). Equation (\ref{eq:projector}) gives an isomorphism between the equivalence classes of the $\mathbb{C}P^{2S}$ model and the set of rank-1 Hermitian projectors. In this formulation, the action functional takes a more compact form which is scaling-invariant
\begin{equation}\label{eq:Action_func_3}
\mathcal{A}(P_k) = \iint_{\mathbb{S}^2}tr\left(\pa P_k\opa P_k\right)\;d\xi_+ d\xi_-
\end{equation}
and its extremum is subject to the constraints (\ref{eq:proj_constraints}). The EL equations corresponding to (\ref{eq:ELeq2}), with constraints (\ref{eq:proj_constraints}), take the simple form \cite{Grundland1}
\begin{equation}\label{eq:ELeq3}
[\pa\opa P_k, P_k] = 0,
\end{equation}
or can be equivalently written as the conservation law
\begin{equation}\label{eq:Conserv_laws}
\pa[\opa P_k, P_k] + \opa[\pa P_k, P_k] = 0.
\end{equation}
For the sake of simplicity, we use the same symbol $0$ for the scalar, vector and zero matrix throughout this paper.

\noindent It was shown \cite{Din1,Din2} that under the assumption that the $\mathbb{C}P^{2S}$ model (\ref{eq:ELeq3}) is defined on the whole Riemann sphere $\Omega = \mathbb{S}^2$ 
and that its action functional (\ref{eq:Action_func_3}) is finite, all rank-1 projectors $P_k(\xi_+, \xi_-)$ can be obtained by acting on the holomorphic (or antiholomorphic) solution $P_0$ with raising (or lowering) operators $\Pi_\pm$. 
In terms of the nonconstant projectors $P_k$, these operators take the form \cite{Goldstein1}
\begin{equation}\label{eq:operators}
P_{k\pm 1} = \Pi_\pm(P_k) =  \frac{\displaystyle (\pa_\pm P_k)P_k(\pa_\mp P_k)}{\displaystyle tr[(\pa_\pm P_k)P_k(\pa_\mp P_k)]}  \qquad \text{for}\quad tr[(\pa_\pm P_k)P_k(\pa_\mp P_k)]\neq 0,
\end{equation}
and are equal to zero when $tr[(\pa_\pm P_k)P_k(\pa_\mp P_k)]=0$. 
Thus, from (\ref{eq:operators}), 
the sequence of solutions in the $\mathbb{C}P^{2S}$ model consists of $(2s+1)$ rank-1 projectors $\{P_0, P_1, ..., P_{2s}\}$. These projectors satisfy the orthogonality and completeness relations \cite{Goldstein3}
\begin{equation}\label{eq:Orthog_Completeness}
P_jP_k = \delta_{jk}P_j, \;(\text{no summation})\qquad \sum_{j = 0}^{2s}P_j = \mathbf{1}_{2s+1}.
\end{equation}
\noindent In what follows in this paper, we will assume that the $\mathbb{C}P^{2S}$ model is defined on the Riemann sphere $\mathbb{S}^2$ and has a finite action functional.

\section{Solutions of the $\mathbb{C}P^{2S}$ sigma model expressed in terms of the Krawtchouk polynomials}
\noindent In this section, we complete the theory
by showing the special class of rank-1 projector $P_k$ of the EL equations (\ref{eq:ELeq3}), which are generated by the initial solution $f_0$ \eqref{eq:Veronese},  can be explicitly expressed in terms of the Krawtchouk polynomials, which is one of the hypergeometric orthogonal polynomials of the Askey scheme \cite{Koekoek1994}. We describe the links between the analytical properties of different solutions obtained through successive applications of a creation and annihilation operator (\ref{eq:recurrence}). Separate classes of holomorphic, antiholomorphic and mixed type solutions are obtained explicitly and expressed in terms of Krawtchouk polynomials. As a result, we find new analytical expressions for the general rank-1 projector solutions of the EL equations (\ref{eq:ELeq3}).

\noindent As explained previously, a holomorphic solution $f_0$ of the $\mathbb{C}P^{2S}$ sigma model (\ref{eq:ELeq2}) can be written as a Veronese sequence of the form (\ref{eq:Veronese}). We note that this Veronese sequence may be expressed in a more compact way 
in terms of the Krawtchouk orthogonal polynomial
\begin{equation}\label{eq:Veronese_2}
(f_0)_j = \sqrt{\binom{2s}{j}} \xi_+^j K_j(k; p, 2s), \qquad \text{for } k=0, \quad 0\leq j\leq 2s,
\end{equation}
\begin{equation}\label{eq:Veronese_2A}
0<p = \frac{\displaystyle \xi_+\xi_-}{\displaystyle 1+\xi_+\xi_-}<1,
\end{equation}
where $(f_0)_j$ is the $j^{\text{th}}$ component of the vector $f_0\in \mathbb{C}^{2s+1}$. Also, we use the convention
\begin{equation}
K_j(0;p,2s)=1.
\end{equation}
We recall that the Krawtchouk polynomial $K_j(k; p, 2s)$ is defined in terms of the hypergeometric function \cite{Koekoek1994}
\begin{equation}\label{eq:Krawtchouk_vs_Hypergeom}
K_j(k)=K_j(k; p, 2s) = \prescript{}{2}F_{1}(-j, -k; -2s; 1/p).
\end{equation} 
Throughout this paper, for the sake of simplicity, we abbreviate some expressions by omitting their arguments. For the Krawtchouk polynomials, we write $K_j$ instead of $K_j(k; p, 2s)$ and $K_j(k\pm1)$ instead of $K_j(k\pm1; p, 2s)$. 

\noindent The subsequent solutions of the $\mathbb{C}P^{2S}$ model (\ref{eq:ELeq3})  can be obtained by acting with the creation or annihilation operators $P_\pm$ through the recurrence relations (\ref{eq:recurrence}). A set of subsequent solutions of the $\mathbb{C}P^{2S}$ model, which consists of $2s+1$ vectors $f_k$, is called a Veronese sequence of solutions if the subsequent solutions are obtained by acting with the creation operators (\ref{eq:recurrence}) on the holomorphic Veronese solution (\ref{eq:Veronese_2}). Thus the set of $(2s+1)$ rank-1 projectors $\{P_0, P_1, ..., P_{2s}\}$ obtained through this procedure satisfies the EL equations (\ref{eq:ELeq3}) and the orthogonality and completeness relations (\ref{eq:Orthog_Completeness}).

\noindent This analysis has opened a new way for constructing and studying this type of solution in terms of the Krawtchouk polynomials. 
The advantage of expressing them in terms of these polynomials lies in the fact that they allow us to find the explicit forms of analytical 
rank-1 solutions, provided that the Euclidean two-dimensional $\mathbb{C}P^{2S}$ model admits a finite action (\ref{eq:Action_func_3}). We obtain the following result.
\begin{theorem}\label{theo}
Let the $\mathbb{C}P^{2S}$ model be defined on the Riemann sphere $\mathbb{S}^2$ and have a finite action functional. Then the Veronese subsequent analytic solutions $f_k$ of the $\mathbb{C}P^{2S}$ model (\ref{eq:ELeq3}) take the form 
\be\label{eq:vector_functions}
(f_k)_{j} = \frac{\displaystyle (2s)!}{\displaystyle (2s-k)!}\left(\frac{\displaystyle -\xi_-}{\displaystyle 1+\xi_+\xi_-}\right)^k \sqrt{\binom{2s}{j}}\xi_+^jK_j(k; p, 2s), \qquad 0\leq k,j\leq2s
\ee
where $K_j(k; p, 2s)$ are the Krawtchouk orthogonal polynomials (\ref{eq:Krawtchouk_vs_Hypergeom}) and where $p$ is given by (\ref{eq:Veronese_2A}).
The rank-1 Hermitian projectors $P_k$ corresponding to the vectors $f_k$ have the form 
\be\label{eq:Projectors_components}
(P_k)_{ij} = \binom{2s}{k}\frac{\displaystyle (\xi_+\xi_-)^{k}}{\displaystyle (1+\xi_+\xi_-)^{2s}}\xi_+^i\xi_-^j\sqrt{\binom{2s}{i}\binom{2s}{j}}K_i(k;p,2s)K_j(k;p,2s).
\ee
\end{theorem}
\textbf{Proof}. The proof for the existence of the Veronese sequence of solutions $f_k$ is straightforward if we use the holomorphic solution $(\ref{eq:Veronese})$ and the recurrence relation (\ref{eq:recurrence}). 
If $f_k$ is given by (\ref{eq:vector_functions}), the orthogonal projector $P_k$ given by (\ref{eq:projector}) can be written as
\begin{align}
(P_k)_{ij} &= \frac{\displaystyle (f_k\otimes f_k^\dagger)_{ij}}{\displaystyle f_k^\dagger f_k} = \frac{\displaystyle \xi_+^i\xi_-^j \sqrt{\binom{2s}{i}\binom{2s}{j}}K_iK_j}{\displaystyle \sum_{q=0}^{2s}\left(\xi_+\xi_-\right)^q \binom{2s}{q}K_q^2}.
\end{align}
By using the orthogonality relation (\ref{eq:ort1}) we find that $(P_k)_{ij}$ is given by (\ref{eq:Projectors_components}).

We now show that, for $0\leq k\leq2s$, the components of the vector functions $(f_k)_j$ given by (\ref{eq:vector_functions}) 
satisfy the recurrence relation (\ref{eq:recurrence})
\be
(f_{k+1})_j-\partial (f_k)_j+(P_k\partial f_k)_j =0.
\ee
First, let us compute the first derivative of $(f_k)_j$ with respect to $\xi_+$. By using relation (\ref{eq:A7}), we obtain
\be
\partial(f_k)_j =\frac{(2s)!}{(2s-k)!}\left(\frac{-\xi_-}{1+\xi_+\xi_-}\right)^k \sqrt{\binom{2s}{j}}\xi_+^{j-1}
\left[(j-k)K_j+\frac{k}{1+\xi_+\xi_-}K_j(k-1)\right].
\ee
Next, by using the explicit form (\ref{eq:Projectors_components}) of $P_k$ and the orthogonality relations (\ref{eq:ort1}) and (\ref{eq:ort2}), 
we get
\begin{eqnarray}
 (P_k\partial f_k)_j=2\frac{(2s)!}{(2s-k)!}\left(\frac{-\xi_-}{1+\xi_+\xi_-}\right)^{k+1} \sqrt{\binom{2s}{j}}\xi_+^{j+1}
(k-s).
\end{eqnarray}
Finally, we can compute
\begin{eqnarray}
(f_{k+1})_j-\partial (f_k)_j+(P_k\partial f_k)_j&=&
\frac{(2s)!}{(2s-k)!}\left(\frac{-\xi_-}{1+\xi_+\xi_-}\right)^{k} \sqrt{\binom{2s}{j}}\xi_+^{j-1}\\
&&\hspace{-3cm}\times\left[-p(2s-k)K_j(k+1)+(k-j+2p(s-k))K_j-k(1-p)K_j(k-1)\right].\nonumber
\end{eqnarray}
The last term vanishes because of (\ref{diff}). 
We have proved that the right-hand side of (\ref{eq:vector_functions}) satisfies the recurrence relation and, for $k=0$, is the holomorphic Veronese sequence.  
Therefore, we conclude that $f_k$ is the Veronese sequence which completes the proof.
$\left.\right.$\hfill$\square$

\noindent Let us remark that another form of the solutions $f_k$ is also given in \cite{Bolton}. 

\noindent By construction, the vectors $f_k$ introduced in the previous theorem are orthogonal to each other.
The orthogonality weights of the Krawtchouk polynomials constitute the binomial distribution as shown in \eqref{eq:ort1}.


\noindent From Theorem \ref{theo}, we are also able to deduce the following useful expressions, 
which are an analogue of the first Frenet formulae \cite{Goldstein3} given in terms of the Krawtchouk polynomials.
\begin{proposition}\label{pr:fr}
The first Frenet formulae associated with the Veronese subsequent solutions of the $\mathbb{C}P^{2S}$ model are given by
 \begin{align}\nonumber
&(P_k \pa P_k)_{il} =-(\pa P_{k-1} P_{k-1})_{il}= \binom{2s}{k}\frac{\displaystyle \sqrt{\binom{2s}{l}\binom{2s}{i}}}{\displaystyle (1+\xi_+\xi_-)^{2s+1}}\ k\xi_+^{k+i-1}\xi_-^{k+l}K_iK_l(k-1),\\\nonumber
&(\opa P_k P_k)_{il} = -( P_{k-1} \opa P_{k-1})_{il} =\binom{2s}{k}\frac{\displaystyle \sqrt{\binom{2s}{l}\binom{2s}{i}}}{\displaystyle (1+\xi_+\xi_-)^{2s+1}}\ k\xi_+^{k+i}\xi_-^{k+l-1}K_i(k-1)K_l,\label{eq:A9}\\\nonumber
&(P_k \opa P_k)_{il} = -(\opa P_{k+1}  P_{k+1})_{il} \nonumber \\
& = \binom{2s}{k}\frac{\displaystyle \sqrt{\binom{2s}{l}\binom{2s}{i}}}{\displaystyle (1+\xi_+\xi_-)^{2s+1}}\ k\xi_+^{k+i}\xi_-^{k+l-1}K_i \nonumber
\left[K_l((l-2s+k)\xi_+\xi_-+l-k)+kK_l(k-1)\right],\\\nonumber
&(\pa P_k P_k)_{il} =-( P_{k+1} \pa P_{k+1})_{il} \nonumber\\
&= \binom{2s}{k}\frac{\displaystyle \sqrt{\binom{2s}{l}\binom{2s}{i}}}{\displaystyle (1+\xi_+\xi_-)^{2s+1}}\ k\xi_+^{k+i-1}\xi_-^{k+l}K_l\left[K_i((i-2s+k)\xi_+\xi_-+i-k)+kK_i(k-1)\right]. \nonumber
\end{align}
\end{proposition}
\textbf{Proof}.
By differentiating the expression (\ref{eq:Projectors_components}) of the projectors $P_k$ with respect to $\xi_+$, one gets
\begin{eqnarray}
(\pa P_k)_{ij} &=& \binom{2s}{k}\sqrt{\binom{2s}{i}\binom{2s}{j}}\frac{\xi_+^{k+i-1}\xi_-^{k+j}}{(1+\xi_+\xi_-)^{2s+1}}
\big[((i-2s)\xi_+\xi_- +i-k(1-\xi_+\xi_-))K_iK_j\nonumber\\
&&\hspace{4cm}+k(K_i(k-1)K_j+K_iK_j(k-1))\big].\label{devP}
\end{eqnarray}
We have used relation (\ref{eq:A7}) to get the previous result.
We can obtain similar relations for  $(\opa P_k)_{ij}$. The first relation of the proposition is proven by computing $\sum_{q=0}^{2s}(P_k)_{iq} (\pa P_k)_{ql}$
and by again using the orthogonality relations of Lemma \ref{lem:ort}. The three other relations are proven similarly.
$\left.\right.$\hfill$\square$

\noindent We now show that the linear spectral problem (LSP) can be expressed in terms of the Krawtchouk orthogonal polynomials $K_k$. 
Indeed, it was shown \cite{Mikhailov,Zakharov} that the $\mathbb{C}P^{2S}$ models with finite action integral (\ref{eq:ELeq3}) are completely integrable. The LSP associated with (\ref{eq:ELeq3}) is given
by $(\lambda \in \mathbb{C})$
\be\label{eq:LSP}
\pa \phi_k = \mathcal{U}(\lambda)\phi_k = \frac{\displaystyle  2}{\displaystyle  1+\lambda}[\pa P_k, P_k]\phi_k,\qquad \opa\phi_k =\mathcal{V}(\lambda)\phi_k =  \frac{\displaystyle  2}{\displaystyle  1-\lambda}[\opa P_k, P_k]\phi_k,
\ee
where $\lambda\in\mathbb{C}$ is the spectral parameter and $P_k$ is the sequence of rank-1 orthogonal projectors (\ref{eq:Orthog_Completeness}) which map onto 
the one-dimensional direction of $f_k$. For all values of $\lambda$, the compatibility condition for equations (\ref{eq:LSP}) corresponds precisely 
to the EL equations (\ref{eq:ELeq3}). By using the results of Proposition \ref{pr:fr}, we find the following
analytical expressions for $\mathcal{U}$ and $\mathcal{V}$
\begin{align}
&(\mathcal{U}(\lambda))_{il} = \frac{\displaystyle 2}{\displaystyle 1+\lambda}\binom{2s}{k}\frac{\displaystyle \sqrt{\binom{2s}{l}\binom{2s}{i}}}{\displaystyle (1+\xi_+\xi_-)^{2s+1}}\ \xi_+^{k+i-1}\xi_-^{k+l}\\\nonumber
&\qquad\qquad\qquad\cdot \left[K_l\left(K_i\left((i-2s+k)\xi_+\xi_-+i-k      \right)+kK_i(k-1)\right)-kK_iK_l(k-1)\right],\\
&(\mathcal{V}(\lambda))_{il} = \frac{\displaystyle 2}{\displaystyle 1-\lambda}\binom{2s}{k}\frac{\displaystyle \sqrt{\binom{2s}{l}\binom{2s}{i}}}{\displaystyle (1+\xi_+\xi_-)^{2s+1}}\ \xi_+^{k+i}\xi_-^{k+l-1}\\\nonumber
&\qquad\qquad\qquad\cdot \left[kK_i(k-1)K_l-K_i\left(K_l\left((l-2s+k)\xi_+\xi_-+l-k      \right)+kK_l(k-1)\right)\right].
\end{align}
An explicit soliton solution which vanishes at complex infinity was found for equations (\ref{eq:ELeq3}) in \cite{Mikhailov,Zakharov}. 
Namely, in the asymptotic case, in which $\phi_k$ tends to the identity matrix $\mathbf{1}_{2s+1}$ as $\lambda$ goes to $\infty$, we have
\begin{align}\nonumber
\phi_k &= \mathbf{1}_{2s+1}+\frac{4\lambda}{(1-\lambda)^2}\sum_{j=0}^{k-1} P_j - \frac{2}{1-\lambda}P_k\\\label{eq:LSP2}
\phi_k^{-1} &= \mathbf{1}_{2s+1}-\frac{4\lambda}{(1+\lambda)^2}\sum_{j=0}^{k-1} P_j-\frac{2}{1+\lambda}P_k,\qquad \lambda = it, \quad t \in\mathbb{R}.
\end{align}
Analytical results for (\ref{eq:LSP2}) have explicitly been carried out for the wavefunction $\phi_k$ but the expressions are rather involved, so we omit them here. 
In accordance with \cite{Sym}, we use the term soliton surfaces, which refers to surfaces associated with an integrable system.
\section{The Clebsch-Gordan decomposition}
Let us now explore certain properties of the $\mathbb{C}P^{2S}$ model. From the Veronese sequence solutions $f_k$ of this model
given in Theorem \ref{theo}, we obtain an explicit expression for the Lagrangian density $L(P_k) = tr(\opa P_k\cdot \pa P_k)$ 
and its action integral $\mathcal{A}(P_k) = \iint_{\mathbb{S}^2}L\;d\xi_+ d\xi_-$ expressed in terms of the Krawtchouk polynomials.
\begin{proposition}
Let the $\mathbb{C}P^{2S}$ model be defined on the Riemann sphere $\mathbb{S}^2$. We recall that 
the rank-1 Hermitian projectors $P_k$, computed in Theorem \ref{theo}, read
\be
(P_k)_{ij} = \binom{2s}{k}\frac{\displaystyle (\xi_+\xi_-)^{k}}{\displaystyle (1+\xi_+\xi_-)^{2s}}\xi_+^i\xi_-^j\sqrt{\binom{2s}{i}\binom{2s}{j}}K_i(k)K_j(k).
\ee
Then the Lagrange density is given by
\be\label{density1}
L(P_k) = \frac{\displaystyle 2(s+2sk-k^2)}{\displaystyle (1+\xi_+\xi_-)^2},
\ee
and the action integral (\ref{eq:Action_func_3}) is finite, positive and takes the form
\begin{equation}\label{eq:Action_func_4}
\mathcal{A}(P_k) =2\pi(s+2sk-k^2)
, \qquad 0\leq k\leq 2s.
\end{equation}
\end{proposition}
\textbf{Proof}. The first derivative with respect to $\xi_+$ of $(P_k)_{ij}$ is given by (\ref{devP}). The complex conjugate provides the 
first derivative with respect to $\xi_-$. Therefore, one gets
\begin{align}
 L&=tr\left(\opa P_k \cdot\pa P_k\right) = \sum_{i, j = 0}^{2s} (\opa P_k)_{ij}(\pa P_k)_{ji}\\
 &= \sum_{i, j = 0}^{2s} \frac{\displaystyle \binom{2s}{k}^2\binom{2s}{i}\binom{2s}{j}}{\displaystyle (1+\xi_+\xi_-)^{4s+2}}(\xi_+\xi_-)^{2k+i+j-1}\\
 &\quad\cdot \left[ K_iK_j\left( j(1+\xi_+\xi_-)
 -2s\xi_+\xi_--k(1-\xi_+\xi_-)\right) +k(K_i(k-1)K_j+K_iK_j(k-1))\right]^2\nonumber
\end{align}
By expanding the square in the last relation and by using the relations in Lemma \ref{lem:ort}, after some algebraic manipulations, 
one proves relation (\ref{density1}).
Thus the $\mathbb{C}P^{2S}$ model is defined from its action integral over the Riemann sphere
\begin{equation}\label{eq:Action_func_3_0}
\mathcal{A}(P_k) = 2(s+2sk-k^2)\iint_{\mathbb{S}^2}\frac{\displaystyle d\xi_+ d\xi_-}{\displaystyle (1+\xi_+\xi_-)^{2}}=2\pi(s+2sk-k^2)
,
\end{equation}
which completes the proof.
$\left.\right.$\hfill$\square$

\noindent Next, suppose that we have constructed a set of rank-1 Hermitian projectors $P_k$ satisfying the EL equations (\ref{eq:ELeq3}) for which the action functional (\ref{eq:Action_func_3}) is required to be finite. It was shown that each second mixed derivative of $P_k$ can be represented as a combination of at most three rank-1 neighbouring projectors \cite{Goldstein4}
\be\label{eq:Mixed_Deriv_Proj}
\pa\opa P_k = \hat{\alpha}_kP_{k-1}+(\hat{\alpha}_k + \check{\alpha}_k)P_k +\check{\alpha}_kP_{k+1}, 
\ee
where the Clebsch-Gordan coefficients $\hat{\alpha}_k$ and $\check{\alpha}_k $ are real-valued functions given by 
\be
\hat{\alpha}_k = tr(\opa P_kP_k \pa P_k) =\hat{\alpha}_k^\dagger ,\qquad \check{\alpha}_k = tr(\pa P_kP_k \opa P_k) =\check{\alpha}_k^\dagger.
\ee
The sum of these coefficients 
\be
\hat{\alpha}_k +\check{\alpha}_k = tr(\pa P_k \opa P_k)
\ee
represents the Lagrangian density (\ref{density1}). Then we have the following
\begin{proposition}
For the subsequent Veronese analytic solutions of the $\mathbb{C}P^{2S}$ model (\ref{eq:ELeq3}), the Clebsch-Gordan coefficients take the simple form
\be\label{eq:Clebsch_Coeff_simple}
\hat{\alpha}_k = \frac{\displaystyle k(2s+1-k)}{\displaystyle (1+\xi_+\xi_-)^{2}}\quad\text{and}\qquad \check{\alpha}_k = \frac{\displaystyle (k+1)(2s-k)}{\displaystyle (1+\xi_+\xi_-)^{2}}.
\ee
\end{proposition}
\textbf{Proof}. From Proposition \ref{pr:fr}, we know the expression of $\left(P_k\pa P_k\right)_{il}$.
Hence the Clebsch-Gordan coefficient $\hat{\alpha}_k$ can be determined explicitly
\begin{align}\nonumber
 \hat{\alpha}_k &= tr(\opa P_kP_k \pa P_k) = \sum_{i,l=0}^{2s}(\opa P_k)_{li}(P_k \pa P_k)_{il}\\\nonumber
 &= \binom{2s}{k}^2\frac{\displaystyle k(\xi_+\xi_-)^{2k-1}}{\displaystyle (1+\xi_+\xi_-)^{4s+2}}\sum_{i,l=0}^{2s}\binom{2s}{l}\binom{2s}{i}\xi_+^{l+i}\xi_-^{i+l}K_iK_l(k-1)\\\nonumber
 &\cdot\left[ \left( i(1+\xi_+\xi_-) -2s\xi_+\xi_- -k(1-\xi_+\xi_-)\right) K_lK_i+k\left( K_l(k-1)K_i+K_lK_i(k-1)\right)\right]\\\nonumber
 &= \binom{2s}{k}^2\frac{\displaystyle k^2(\xi_+\xi_-)^{2k-1}}{\displaystyle (1+\xi_+\xi_-)^{4s+2}}\sum_{i,l}\binom{2s}{l}\binom{2s}{i}\xi_+^{l+i}\xi_-^{l+i}K_l^2(k-1)K_i^2.
\end{align}
By using the orthogonality relation of the Krawtchouk polynomials (\ref{eq:ort1}), we prove the first relation of (\ref{eq:Clebsch_Coeff_simple}).
Bearing in mind that
\be\label{eq:temp2}
\check{\alpha}_k = tr(\pa P_kP_k \opa P_k) = tr(\opa P_k\pa P_k) -\hat{\alpha}_k
\ee
and making use of (\ref{density1}), we get the 
second relation of (\ref{eq:Clebsch_Coeff_simple}) which completes the proof.
$\left.\right.$\hfill$\square$

This computation gives an explicit form of the Clebsch-Gordan coefficients $\hat{\alpha}_k$ and $\check{\alpha}_k$ 
in terms of the complex independent variables $(\xi_+,\xi_-)\in\mathbb{C}$.
It follows from (\ref{eq:Mixed_Deriv_Proj}) and (\ref{eq:Clebsch_Coeff_simple}) that the second mixed derivative can be represented as a linear combination of three rank-1 neighbouring projectors, namely
\begin{equation}\label{eq:Mixed_Deriv_Proj_2}
\pa\opa P_k = \frac{\displaystyle 1}{\displaystyle (1+\xi_+\xi_-)^{2}} \left[k(2s-k-1)P_{k-1}+2(s+2sk-k^2)P_k+(k+1)(2s-k)P_{k+1}\right].
\end{equation}
From (\ref{eq:Orthog_Completeness}) and (\ref{eq:Mixed_Deriv_Proj_2}), we show that $P_k$ also satisfies
\be
tr(P_k\pa\opa P_k) = \frac{\displaystyle 2(s+2sk-k^2)}{\displaystyle (1+\xi_+\xi_-)^{2}}.
\ee
In \cite{Goldstein4}, it was demonstrated that
\begin{equation}
\opa P_k\pa P_k =  \hat{\alpha}_kP_{k-1}+\check{\alpha}_kP_k\quad\text{and}\qquad \pa P_k\opa P_k= \hat{\alpha}_k P_k +\check{\alpha}_kP_{k+1}.
\end{equation}
From the explicit form of the Clebsch-Gordan coefficients (\ref{eq:Clebsch_Coeff_simple}), these expressions become
\begin{align}
\opa P_k\pa P_k &= \frac{\displaystyle 1}{\displaystyle (1+\xi_+\xi_-)^{2}}\left[k(2s-k+1)P_{k-1}+(k+1)(2s-k)P_k\right],\label{eq:dd0}\\
\pa P_k\opa P_k &= \frac{\displaystyle 1}{\displaystyle (1+\xi_+\xi_-)^{2}}\left[(k+1)(2s-k)P_{k+1}+k(2s-k+1)P_k\right].\label{eq:dd}
\end{align}

\section{The $\mathfrak{su}(2)$ spin-$s$ representation associated with the Veronese sequence of solutions of the $\mathbb{C}P^{2S}$ model}
The objective of this section is to demonstrate a connection between the Euclidean sigma model (\ref{eq:ELeq3}) 
in two dimensions and the spin-s $\mathfrak{su}(2)$ representation. The spin matrix $S^z$ is defined as a 
linear combination of the $(2s+1)$ rank-1 Hermitian projectors $P_k$ by \cite{Goldstein2}
\be\label{eq:Spin_matrix_linear_comb}
S^z=\sum_{k=0}^{2s}(k-s)P_k.
\ee
The eigenvalues of the generator $S^z$ are $\{-s,-s+1,\dots,s\}$. They are either integer (for odd $2s+1$) or half-integer (for even $2s+1$) values.
Under these assumptions, we have the following.
\begin{proposition}\label{eq:prop_5_1}
The spin matrix $S^z$ is given by the tridiagonal matrix
\begin{multline}\label{eq:Spin_matrix_tridiagonal}
(S^z)_{ij} = \delta_{ij}\left(\frac{1-\xi_+\xi_-}{1+\xi_+\xi_-}\right)(i-s) - \delta_{i-1,j}\left(\frac{\xi_+}{1+\xi_+\xi_-}\right)\sqrt{i(2s+1-i)}\\
- \delta_{i,j-1}\left(\frac{\xi_-}{1+\xi_+\xi_-}\right)\sqrt{j(2s-j+1)},\qquad 0\leq i, j \leq 2s.
\end{multline}
\end{proposition}
\textbf{Proof}. 
By using the explicit form of the projector $P_k$ found in Theroem \ref{theo}, one gets
\begin{align}
(S^z)_{ij}&=\frac{\xi_+^i\xi_-^j}{(1+\xi_+\xi_-)^{2s}}\sqrt{\binom{2s}{i}\binom{2s}{j}}\sum_{k=0}^{2s}(k-s)\binom{2s}{k}(\xi_+\xi_-)^kK_i(k)K_j(k). \label{eq:temp5}
\end{align}
By using the orthogonality properties (\ref{eq:ortd1})-(\ref{eq:ortd3}), we obtain
\begin{align}
({S^z})_{ij}&= \sqrt{\binom{2s}{i}\binom{2s}{j}}
\left[\delta_{ij}\frac{2s\xi_+\xi_-^2+i-j\xi_+\xi_-^2}{\xi_-(1+\xi_+\xi_-)}\frac{j!(2s-j)!}{(2s)!}
-\delta_{i-1,j} \frac{i!\xi_+}{1+\xi_+\xi_-}\frac{(2s-i+1)!}{(2s)!}\right.
\\&\left.\qquad\qquad\qquad\qquad
-\delta_{i,j-1} \frac{\xi_- j!(2s-j+1)}{(1+\xi_+\xi_-)(2s)!} - \delta_{ij}\frac{sj!(2s-j)!}{(2s)!}
\right].
\end{align}
After algebraic computations, the spin matrix $S^z$ takes the form (\ref{eq:Spin_matrix_tridiagonal}).
$\left.\right.$\hfill$\square$

\noindent We recall that the generators $S^z$ and $S^\pm$ of the $\mathfrak{su}(2)$ Lie algebra satisfy the commutation relations
\be\label{eq:Commutation_generators}
(i) \quad [S^z, S^\pm]= \pm S^\pm,\qquad (ii)\quad [S^+, S^-] = 2S^z.
\ee
Due to the eigenvalues of the spin matrix $S^z$ (\ref{eq:Spin_matrix_linear_comb}), 
we know that this matrix can be seen as the Cartan element of $\mathfrak{su}(2)$ in the spin-s representation. 
We now compute the corresponding $S^+$ and $S^-$ in this representation. The usual spin-$s$ representation of $\mathfrak{su}(2)$ is identified with \cite{Schiff,Merzbacher}
\begin{align}\label{eq:Spin_rep_Schiff}
(\sigma^z)_{ij} &= (s-i)\delta_{ij},\\
(\sigma^+)_{ij} &= \sqrt{(2s-j+1)j}\delta_{i,j-1}, \qquad\qquad 0\leq i, j \leq 2s\\
(\sigma^-)_{ij} &=  \sqrt{(2s-i+1)i}\delta_{i-1,j}.
\end{align}
Due to Proposition \ref{eq:prop_5_1}, we note that we can decompose the spin matrix $S^z$ as a 
linear combination of matrices $\sigma^z$ and $\sigma^\pm$, namely
\be\label{eq:decomp_S_z}
S^z = \frac{\displaystyle 1}{\displaystyle 1+\xi_+\xi_-}\left((\xi_+\xi_- - 1)\sigma^z - \xi_+\sigma^- -\xi_- \sigma^+\right).
\ee
Hence we have
\begin{proposition}\label{eq:prop_5_2}
Let us define the spin matrices by
\begin{align}
S^+ &= \frac{\displaystyle 1}{\displaystyle 1+\xi_+\xi_-}\left(2\xi_-\sigma^z - \sigma^- +\xi_-^2 \sigma^+\right),\\\label{eq:Spin_rep_SchiffS}
S^- &= \frac{\displaystyle 1}{\displaystyle 1+\xi_+\xi_-}\left(2\xi_+\sigma^z + \xi_+^2\sigma^- -\sigma^+\right).
\end{align}
Then the matrix $S^z$ given by (\ref{eq:decomp_S_z}) and the two previous matrices $S^+$ and $S^-$ satisfy 
the $\mathfrak{su}(2)$ commutation relations (\ref{eq:Commutation_generators}).

\noindent Moreover, we have the following useful recurrence properties for $S^zf_k$ and $S^\pm f_k$ when $f_k$ is an analytic vector given by (\ref{eq:vector_functions})
\begin{align}\label{eq:proprietes1}
&(i)\quad S^+f_k =\begin{cases} -(1+\xi_+\xi_-)f_{k+1},& \text{for }\; 0 \leq k\leq 2s-1, \\
 0, & \text{for } k=2s,
\end{cases}\\\label{eq:proprietes2}
&(ii)\quad S^-f_k = \frac{\displaystyle 1}{\displaystyle 1+\xi_+\xi_-}k(k-1-2s)f_{k-1}, \qquad \text{for }\; 0 \leq k\leq 2s,\\\label{eq:proprietes}
&(iii)\quad S^zf_k = (k-s)f_k,\qquad\qquad\qquad\quad\;\, \quad\text{for }\; 0 \leq k\leq 2s,\\
&(iv)\quad S^z(S^\pm f_k)=(k\pm 1 -s)(S^\pm f_k).
\end{align}
So $S^\pm$ are the creation and annihilation operators for $f_k$.
\end{proposition}
\textbf{Proof}. By using the fact that $\sigma^z$ and $\sigma^\pm$ satisfy the $\mathfrak{su}(2)$ commutation relations, we prove that $S^z$ and $S^\pm$ 
also satisfy these relations. 
The property (\ref{eq:proprietes}) follows directly from the definition (\ref{eq:Spin_matrix_linear_comb}) of $S^z$ 
and of the definition (\ref{eq:projector}) of $P_k$. 
The action of the matrix $S^+$ on the vector $f_k$ is given by 
\begin{align}
(S^+f_k)_j &= \frac{1}{1+\xi_+\xi_-}\left[\xi_-^2\sqrt{(2s-j)(j+1)}(f_{k})_{j+1}+2(s-j)\xi_-(f_{k})_{j} -\sqrt{(2s-j+1)j}(f_{k})_{j-1}\right].
\end{align}
By using the explicit form (\ref{eq:vector_functions}) of $f_k$ and identities on binomials we get
\begin{align}
(S^+f_k)_j &= \frac{-(2s)!(\xi_+)^j(-\xi_-)^{k+1}\sqrt{\binom{2s}{j}}}{(2s-k)!(1+\xi_+\xi_-)^{k+1}}
\left[\xi_-\xi_+(2s-j)K_{j+1}(k)+ 2(s-j)K_j(k)-\frac{j}{\xi_+\xi_-}K_{j-1}(k)\right].
  \end{align}
By using relation (\ref{eq:d4}), we obtain for $0\leq k <2s$
\begin{eqnarray}
 (S^+f_k)_j &= -\frac{(2s)!(\xi_+)^j(-\xi_-)^{k+1}\sqrt{\binom{2s}{j}}}{(2s-k-1)!(1+\xi_+\xi_-)^{k}}K_j(k+1).
\end{eqnarray}
By using the explicit form for $f_{k+1}$, we prove (\ref{eq:proprietes1}). The case $k=2s$ is obtained by remarking that the right-hand side of  (\ref{eq:d4})
vanishes in this case.
We perform similar computations to prove (\ref{eq:proprietes2}) which concludes the proof.
$\left.\right.$\hfill $\square$

\noindent  Let us emphasize that relation (\ref{eq:proprietes1}) allows us to recursively construct the sequence $f_k$ starting from $f_0$.
It provides an equivalent way to construct the set of solutions $f_k$ simpler than the ones given by the recurrence relation (\ref{eq:recurrence}). 
We have also similar recurrence relations for the projectors $P_k$
\begin{equation}
 P_{k+1}=\frac{\displaystyle S^+ P_k S^-}{\displaystyle tr(S^+ P_k S^-)} \quad\text{and} \qquad  P_{k-1}=\frac{\displaystyle S^- P_k S^+}{\displaystyle tr(S^- P_k S^+)} \ ,
\end{equation}
where $tr(S^\pm P_k S^\mp)\neq 0$.


\section{Geometrical aspects of surfaces associated with the $\mathbb{C}P^{2S}$ model}
\noindent Let us now study certain geometrical properties of 2-dimensional (2D) surfaces associated with the $\mathbb{C}P^{2S}$ model. We show that these surfaces are immersed in the $\mathfrak{su}(2s+1)$ algebra and may be expressed in terms of the Krawtchouk orthogonal polynomials. These geometrical properties include, among others, the Gaussian and mean curvatures, the topological charge, the Willmore functional and the Euler-Poincar\'e character \cite{Carmo,Nomizu}.

\noindent Under the assumption that the $\mathbb{C}P^{2S}$ model is defined on the Riemann sphere $\mathbb{S}^2$ and that the associated action functional (\ref{eq:Action_func_4}) of this model is finite, we can show that these surfaces are conformally parametrised. The proof is similar to that given in \cite{Goldstein1}. The further advantage of using the $\mathbb{C}P^{2S}$ model in this context lies in the fact that it allows us to provide an explicit parametrisation in terms of the Krawtchouk polynomials instead of the formalism of connected rank-1 projectors. This approach has opened a new way for constructing and investigating 2D-surfaces immersed in multidimensional Euclidean spaces $\mathbb{R}^{4s(s+1)}$.

\noindent Let us now present certain geometrical aspects of 2D-surfaces immersed in the $\mathfrak{su}(2s+1)$ algebras. For a given set of $(2s+1)$ rank-1 projector solutions of the EL equations (\ref{eq:ELeq3}), the generalized Weierstrass formula for immersion (GWFI) of 2D-surfaces is defined by a contour integral \cite{Konopelchenko}
\be\label{eq:GWFI}
X_k(\xi_+, \xi_-) = i\int_{\gamma_k}\left(-[\pa P_k, P_k]d\xi_++[\opa P_k, P_k]d\xi_-\right)\in\mathfrak{su}(2s+1)\simeq \mathbb{R}^{4s(s+1)},
\ee
which is independent of the path of integration $\gamma_k\subset\mathbb{C}$. The conservation laws (\ref{eq:Conserv_laws}) ensure that the contour integral is locally independent of the trajectory. For the surfaces corresponding to the rank-1 projectors $P_k$, the integration of (\ref{eq:GWFI}) can be performed explicitly (since $d(dX_k)=0$) with the result \cite{Grundland2}
\be\label{eq:GWFI_primitive}
X_k = -i\left(P_k+2\sum_{j=0}^{k-1}P_j\right)+i\left(\frac{\displaystyle 1+2k}{\displaystyle 1+2s}\right)\mathbf{1}_{2s+1}, \qquad 0\leq k\leq 2s.
\ee
Note that for each $k$, the projectors $P_j$ satisfy the eigenvalue equations
\be
\left(X_k - i\lambda_k\mathbf{1}_{2s+1}\right)P_j = 0
\ee
with the eigenvalues 
\be
\lambda_k = \left\{\begin{matrix}\frac{\displaystyle 2(k-2s)-1}{\displaystyle 1+2s}\quad \text{for } \; j<k \\\\
\frac{\displaystyle 2(k-s)}{\displaystyle 1+2s}\qquad\; \text{for } \; j=k\\\\
\frac{\displaystyle 1+2k}{\displaystyle 1+2s}\qquad\;\;\, \text{for } \; j>k
\end{matrix}\right..
\ee
The immersion functions $X_k$ span a Cartan subalgebra of $\mathfrak{su}(2s+1)$
\be
[X_k, X_j]=0, \qquad 0 \leq k, j \leq 2s
\ee
and satisfy the algebraic conditions given in \cite{Goldstein5}. For mixed solutions $P_k$, $1\leq k\leq 2s$, we get a cubic matrix equation
\be
\left[X_k - i \frac{\displaystyle 1+2k}{\displaystyle 1+2s}\mathbf{1}_{2s+1}\right]\left[ X_k - i\frac{\displaystyle 2(k-s)}{\displaystyle 1+2s}\mathbf{1}_{2s+1}\right]
\left[X_k - i\frac{\displaystyle 2(k-2s)-1}{\displaystyle 1+2s}\mathbf{1}_{2s+1} \right] = 0
\ee
while for holomorphic $(k=0)$ and antiholomorphic $(k = 2s)$ solutions of the $\mathbb{C}P^{2S}$ equation (\ref{eq:ELeq3}), the minimal polynomial for the real-valued matrix function $X_k$ is quadratic. Namely, we have
\be
\left[X_0 -  \frac{\displaystyle i}{\displaystyle 1+2s}\mathbf{1}_{2s+1}\right]\left[ X_0 + \frac{\displaystyle 2is}{\displaystyle 1+2s}\mathbf{1}_{2s+1}\right]=0, \qquad k=0,
\ee
and
\be
\left[X_{2s} +  \frac{\displaystyle i}{\displaystyle 1+2s}\mathbf{1}_{2s+1}\right]\left[ X_{2s} - \frac{\displaystyle 2is}{\displaystyle 1+2s}\mathbf{1}_{2s+1}\right]=0, \qquad k=2s.
\ee
For the sake of uniformity, the inner product $X_k$ is defined by \cite{Helgason}
\be\label{eq:scalar}
(A, B) = -\frac{1}{2}tr(A\cdot B), \qquad\qquad \text{for any }\;A, B \in\mathfrak{su}(2s+1),
\ee
instead of the Killing form. In view of the analytical form (\ref{eq:GWFI_primitive}) of the 2D-surfaces $X_k$ given in terms of the projectors $P_k$, which are expressed through the formula (\ref{eq:Projectors_components}), we can determine that the 2D-surfaces associated with the $\mathbb{C}P^{2S}$ model are submanifolds of the compact sphere with radius
\be\label{radius1}
(X_k, X_k) = -\frac{\displaystyle  1}{\displaystyle  2}tr(X_k)^2 = \frac{\displaystyle 1}{\displaystyle 2}\left(\frac{\displaystyle (1+2k)(2(2s-k)-1)}{\displaystyle 1+2s}-1\right)
\ee
immersed in $\mathbb{R}^{4s(s+1)}\simeq \mathfrak{su}(2s+1)$.

\noindent The projectors $P_k$ fulfill the completeness relation (\ref{eq:Orthog_Completeness}) which implies in turn that the immersion functions $X_k$ satisfy the linear relation \cite{Goldstein5}
\be
\sum_{k=0}^{2s}(-1)^kX_k = 0.
\ee
Once we have determined the immersion functions (\ref{eq:GWFI_primitive}) of the 2D-surfaces, we can describe their metric and curvatures properties.
From (\ref{eq:GWFI}) and Proposition \ref{pr:fr}, the complex tangent vectors are obviously
\begin{align}\nonumber
&(\pa X_k)_{rl} = -i\left( [\pa P_k, P_k] \right)_{rl} = -i\binom{2s}{k}\frac{\sqrt{\binom{2s}{l}\binom{2s}{r}}}{(1+\xi_+\xi_-)^{2s+1}}\xi_+^{k+r-1}\xi_-^{k+l}\\\nonumber
&\qquad\cdot\left[K_l\left(K_r((r-2s+k)  \xi_+\xi_-+r-k        )+kK_r(k-1)\right)-kK_rK_l(k-1)\right],\\\nonumber
&(\opa X_k)_{rl} = i\left( [\opa P_k, P_k] \right)_{rl} = i\binom{2s}{k}\frac{\sqrt{\binom{2s}{l}\binom{2s}{r}}}{(1+\xi_+\xi_-)^{2s+1}}\xi_+^{k+r}\xi_-^{k+l-1}\\
&\qquad\cdot\left[kK_r(k-1)K_l - K_r\left(K_l((l-2s+k)  \xi_+\xi_-+l-k        )+kK_l(k-1)\right)\right]\label{tangentthing}.
\end{align}
Let $g_k$ be the metric tensor associated with the surface $X_k$. Its components are indicated with indices outside of parentheses in order to 
distinguish them from the index of the surface $X_k$. The diagonal elements of the metric tensor are zero (i.e. $(g_k)_{11}=(g_k)_{22}=0$). 
This property follows from the vanishing of $tr(\pa P_k\cdot \pa P_k)$ or $tr(\opa P_k\cdot \opa P_k)$, as proven in \cite{Goldstein1}. 
In view of relation (\ref{eq:dd}), the non-zero off-diagonal elements are
\begin{align}\label{g12element1}
(g_k)_{12} & = (g_k)_{21}= -\frac{1}{2}tr(\pa X_k\cdot \opa X_k) = \frac{1}{2}tr(\pa P_k\cdot \opa P_k) = \frac{\displaystyle s(2k+1)-k^2}{\displaystyle (1+\xi_+\xi_-)^2}, \qquad 0\leq k\leq 2s.
\end{align}
The area of the 2D-surfaces given by $\iint_{\mathbb{S}^2}(g_k)_{12} d\xi_+ d\xi_-$ are equal to the action integral \eqref{eq:Action_func_4}. 
Thus the first fundamental forms reduce to
\be
I_k = tr(\pa P_k\cdot \opa P_k)d\xi_+d\xi_- = 2\ \frac{\displaystyle s(2k+1)-k^2}{\displaystyle (1+\xi_+\xi_-)^2}d\xi_+d\xi_-
\ee
The non-zero Christoffel symbols of the second type are
\begin{align}\nonumber
(\Gamma_k)_{11}^1 &= \pa \ln(g_k)_{12} = - \frac{\displaystyle 2}{\displaystyle 1+\xi_+\xi_-}\xi_-,\\
(\Gamma_k)_{22}^2 &= \opa \ln(g_k)_{12} = - \frac{\displaystyle 2}{\displaystyle 1+\xi_+\xi_-}\xi_+.
\end{align}
We observe that the 2D-surfaces $X_k$ are torsion-free, i.e. $(T_k)^a_{bc} = (\Gamma_k)^a_{bc}-(\Gamma_k)^a_{cb} = 0$.
Hence the second fundamental forms
\begin{eqnarray}
 II_k& =& \left(\pa^2X_k - (\Gamma_k)_{11}^1\pa X_k\right)d\xi_+^2+2\pa\opa X_k d\xi_+d\xi_- + \left(\opa^2X_k - (\Gamma_k)_{22}^2\opa X_k\right)d\xi_-^2 \ \ \ \ \ \\\
 &=&-tr(\pa P_k \opa P_k) \pa \left( \frac{[\pa P_k,P_k]}{tr(\pa P_k \opa P_k)} \right) d\xi_+^2 +2i [\opa P_k, \pa P_k]d\xi_+d\xi_-\nonumber\\
 &&-tr(\pa P_k \opa P_k) \opa \left( \frac{[\opa P_k,P_k]}{tr(\pa P_k \opa P_k)} \right) d\xi_-^2
\end{eqnarray}
are easy to find but the expressions are rather involved, so we omit them here. The first and second fundamental forms allow us to formulate the following.
\begin{proposition}
The Gaussian curvatures
\be\label{gaussiancurvature1}
\mathcal{K}_k = \frac{\displaystyle -2\pa\opa\ln|tr(\pa P_k\cdot\opa P_k)|}{\displaystyle tr(\pa P_k\cdot\opa P_k)}.
\ee
of 2D-surfaces associated with the Veronese subsequent analytic solutions $f_k$ (\ref{eq:vector_functions}) of the $\mathbb{C}P^{2S}$ model (\ref{eq:ELeq3}) have the following constant positive values
\be\label{gaussiancurvature2}
\mathcal{K}_k = \frac{\displaystyle 2}{\displaystyle 2sk+s-k^2}\ ,\quad 0\leq k\leq 2s.
\ee
\end{proposition}
\textbf{Proof}. Using the complex tangent vectors (\ref{tangentthing}) and the relation (\ref{eq:dd}), we can compute explicit analytic 
expressions for the Gaussian curvatures. In fact, from the formula (\ref{gaussiancurvature1}), we get
\be
\mathcal{K}_k = \frac{\displaystyle -2\pa\opa\ln|2(2sk+s-k^2)(1+\xi_+\xi_-)^{-2}|}{\displaystyle 2(2sk+s-k^2)(1+\xi_+\xi_-)^{-2}}.
\ee
which, after simplification, gives the expression (\ref{gaussiancurvature2}).
$\left.\right.$\hfill$\square$
\begin{proposition}
The mean curvature vectors written in the matrix form
\be
\mathcal{H}_k = \frac{\displaystyle -4i[\pa P_k, \opa P_k]}{\displaystyle tr(\pa P_k\cdot\opa P_k)},
\ee
of the 2D-surfaces $X_k$ associated with the Veronese subsequent analytic solutions $f_k$ (\ref{eq:vector_functions}) of the $\mathbb{C}P^{2S}$ model (\ref{eq:ELeq3}) are expressible in terms of the Krawtchouk polynomials only.
\be\label{meancurvature1}
\begin{split}
&(\mathcal{H}_k)_{jl} = -2i\binom{2s}{k}\frac{\displaystyle \sqrt{\binom{2s}{l}\binom{2s}{j}}}{\displaystyle (1+\xi_+\xi_-)^{2s-1}}\ \ \frac{\displaystyle \xi_+^{k+l-1}\xi_-^{k+l-1}}{\displaystyle 2sk+s-k^2}\cdot\\
& \Big{\{}K_lK_j\big{[}\alpha_2(\xi_+\xi_-)^2+\alpha_1\xi_+\xi_-+\alpha_0\big{]}+kK_lK_j(k-1)\big{[}(l-2s+k)\xi_+\xi_-+l-k\big{]}\\ &+kK_jK_l(k-1)\big{[}(j-2s+k)\xi_+\xi_-+j-k\big{]} \Big{\}},
\end{split}
\ee
where
\begin{displaymath}
\alpha_2=(j-2s+k)(l-2s+k),\quad \alpha_1=2\big{(}(j-s)(l-s)-(k-s)(k-s-1)\big{)},\quad \alpha_0=(j-k)(l-k).
\end{displaymath}
The mean curvature vectors $\mathcal{H}_k$ satisfy the following conditions
\be
tr(\mathcal{H}_k) = 0,\qquad (\mathcal{H}_k, \pa X_k) = 0, \qquad (\mathcal{H}_k, \opa X_k) = 0.
\ee
\end{proposition}
\textbf{Proof}. The proof is straightforward if we use relations given by (\ref{eq:dd0})-(\ref{eq:dd}) and relations given in the Appendix.
$\left.\right.$\hfill$\square$

\noindent At this point, we can explore certain global geometrical characteristics of the soliton surfaces $X_k$. 
In particular, we assume that these surfaces are compact, oriented and connected. To compute them we perform an integration over the whole Riemann sphere $\mathbb{S}^2$. Under these circumstances, the following holds.
\begin{proposition}
The Willmore functionals
\be
\mathcal{W}_k = \iint_{\mathbb{S}^2} tr([\pa P_k, \opa P_k])^2 d\xi_+d\xi_-
\ee
of 2D-surfaces associated with Veronese subsequent analytical solutions $f_k$ (\ref{eq:vector_functions}) of the $\mathbb{C}P^{2S}$ model (\ref{eq:ELeq3}) have constant positive values
\be\label{willmore1}
\mathcal{W}_k = \frac{\displaystyle 2\pi}{\displaystyle 3}\big{[}4s^2(k^2+k+1)-2ks(2k^2+k+3)+k^2(k^2+3)\big{]}.
\ee
\end{proposition}
\textbf{Proof}. The proof is straightforward if we use (\ref{eq:dd0})-(\ref{eq:dd}) and relations of the Appendix.
$\left.\right.$\hfill$\square$
\begin{proposition}
For any value of the Veronese subsequent analytic solutions $f_k$ (\ref{eq:vector_functions}) of the $\mathbb{C}P^{2S}$ model (\ref{eq:ELeq3}), the integral over the whole Riemann sphere $\mathbb{S}^2$ of the topological charge densities of the 2D-surfaces $X_k$
\be\label{topological1}
Q_k = \frac{\displaystyle  1}{\displaystyle  \pi}\iint_{\mathbb{S}^2}\pa\opa\ln(f_k^\dagger f_k)d\xi_+d\xi_- ,
\ee
are integers
\be\label{topological2}
Q_k =2(s-k)
\ee
\end{proposition}
\textbf{Proof}. The proof follows immediately from (\ref{eq:vector_functions}) since the relation
\begin{equation}
 f_k^\dagger f_k=\frac{\displaystyle (2s)! k!}{\displaystyle (2s-k)!}( 1+\xi_+\xi_-)^{2(s-k)}
\end{equation}
holds.
$\left.\right.$\hfill$\square$

\noindent Note that the values of the topological charges are distinguished between a one-instanton state $Q_0=2s$ in the $\mathbb{C}P^{2S}$ model and an anti-instanton state $Q_{2s}=-2s$. This fact produces the same winding over the target sphere $\mathbb{S}^2$ but in the opposite direction. This result coincides with the values of the topological charges obtained earlier for low-dimensional sigma models \cite{Goldstein1}.
\begin{proposition}
For any value of the Veronese subsequent analytic solutions $f_k$ (\ref{eq:vector_functions}) of the $\mathbb{C}P^{2S}$ model (\ref{eq:ELeq3}), the Euler-Poincar\'e characters of the 2D-surfaces $X_k$ 
\be\label{eulerpoincare1}
\Delta_k = -\frac{\displaystyle 1}{\displaystyle \pi}\iint_{\mathbb{S}^2}\pa\opa \ln|tr(\pa P_k\cdot\opa P_k)|d\xi_+d\xi_-.
\ee
are the integer
\be\label{eulerpoincare2}
\Delta_k = 2,
\ee
for all $k$ such that $0\leq k\leq 2s$.
\end{proposition}
\textbf{Proof}. The proof follows directly from (\ref{eq:dd}).
$\left.\right.$\hfill$\square$

\noindent Note that all surfaces possess the same value of the Euler-Poincar\'e character equal to $2$ and all Gaussian curvatures $\mathcal{K}_k$ are positive. This means that all 2D-surfaces associated with the $\mathbb{C}P^{2S}$ model are homeomorphic to spheres with radius given by (\ref{radius1}).

\section{Concluding remarks}
The approach elaborated in this paper based on the hypergeometric orthogonal polynomials for studying the $\mathbb{C}P^{2S}$ sigma model has proven to be a very useful and suitable tool for investigating 
the main features of numerous problems that require the solving of this model. The task of obtaining an increasing number of solutions of this model is related to the construction of the consecutive projectors $P_k$
and the associated immersion functions $X_k$ expressed in terms of the Krawtchouk polynomials. Their main advantage appears when these polynomials make it possible to construct a regular algorithm for finding
the Veronese subsequent analytical solutions of the $\mathbb{C}P^{2S}$ sigma model without referring to any additional considerations but proceeding only from the given model. A $\mathfrak{su}(2)$ spin-s 
representation in connection with the Krawtchouk polynomials associated with this model has been explored in detail. Based on this connection, we can derive solutions of the $\mathbb{C}P^{2S}$ model through algebraic recurrence relations which are simpler than the known analytic relations. Some geometrical aspects of soliton surfaces have also been described in terms of the Krawtchouk polynomials.
This analysis has opened a new way for computing the associated metric, the first and second fundamental form, the Gaussian and mean curvatures, the Willmore functionals, 
the topological charge and the Euler-Poincar\'e characters. 

\noindent In the next stage of this research, it would be worthwhile to extend this investigation to the case of the Grassmannian sigma models which take values in the homogeneous spaces
\begin{equation}
 G(m,n)=SU(m+n)/S( U(m) \times U(n) )\ .
\end{equation}
These models are a generalisation of the model considered in this paper. Both models possess many common properties like an infinite number of conserved quantities and infinite dimensional 
dynamical symmetries which generate the affine Kac--Moody algebra. They admit Hamiltonian structures and complete integrability with well-formulated linear spectral problems.
The investigation of the soliton surfaces for Grassmannian sigma model can lead to different classes and much more diverse types of surfaces. The question of the diversity and complexity of the 
associated surfaces still remains open for further research and should be answered in further work.

\noindent {\bf Acknowledgements}\\
This research was supported by the NSERC operating grant of one of the authors (A.M.G.).\\
N.C. is indebted to the Centre de Recherches Math\'ematiques (CRM) for the opportunity to hold a CRM-Simons professorship.



%

{}


%
\appendix
\section{Properties of the Krawtchouk polynomials}
In this appendix, we recall and prove useful properties of the Krawtchouk polynomials and of the projectors $P_k$ (\ref{eq:Projectors_components})
According to R. Koekoek \cite{Koekoek1994}, the properties of the Krawtchouk polynomials follow directly from their definition 
\be
K_j(k; p, 2s) = {}_2 F_{1}\left(-j, -k; -2s; 1/p\right), \qquad 0\leq j,k\leq 2s,\label{eq:A3}
\ee
where we recall that $0<p=\frac{\xi_+\xi_-}{1+\xi_+\xi_-}<1$.
In view of the formula (1.10.6) from \cite{Koekoek1994}, the Krawtchouk polynomial forward shift operator is
\be
K_j(k+1; p, 2s) - K_j(k; p, 2s) = -\, \frac{\displaystyle j}{\displaystyle 2sp}K_{j-1}(k;p,2s-1).\label{eq:A6}
\ee

\begin{lemme}
 The first derivative with respect to $\xi_\pm$ of the Krawtchouk polynomials are given by 
\begin{align}
\pa K_j(k;p,2s) &= \frac{\displaystyle -k}{\displaystyle\xi_+(1+\xi_+\xi_-)}(K_j(k;p,2s)-K_{j}(k-1;p,2s)),\label{eq:A7}\\
\opa K_j(k;p,2s) &= \frac{\displaystyle -k}{\displaystyle \xi_-(1+\xi_+\xi_-)}(K_j(k;p,2s)-K_{j}(k-1;p,2s)). \label{eq:AA8}
\end{align}
\end{lemme}
\textbf{Proof}.
We recall the formula for the derivative of the hypergeometric function
\be
\frac{\partial}{\partial x} \left( {}_{2}F_{1}(a, b; c; x)\right) = \frac{\displaystyle ab}{\displaystyle c}\  {}_{2}F_{1}(a+1, b+1; c+1; x).
\ee
Hence, from the definition (\ref{eq:A3}), we can evaluate the first derivatives with respect to $\xi_+$ of the Krawtchouk polynomial
\begin{align}\nonumber
\partial K_j\left(k; \frac{\xi_+\xi_-}{(1+\xi_+\xi_-)}, 2s\right) &= \partial \left( {}_{2}F_{1}\left(-j, -k; -2s; \frac{1+\xi_+\xi_-}{\xi_+\xi_-}\right)\right)\\\nonumber
 &= -\frac{\xi_-}{(\xi_+\xi_-)^2}\left(\frac{-jk}{2s}\right)\prescript{}{2}F_{1}\left(-j+1, -k+1; -2s+1; \frac{1+\xi_+\xi_-}{\xi_+\xi_-}\right)\\
 &=\frac{1}{\xi_+^2\xi_-}\frac{jk}{2s}K_{j-1}(k-1; p, 2s-1).
\end{align}
In view of (\ref{eq:A6}), we obtain the expression (\ref{eq:A7}).
Similarly, we obtain the relation (\ref{eq:AA8}).
$\left.\right.$\hfill$\square$

\begin{lemme}\label{lem:ort}
The following orthogonality relations hold
 \begin{align}
&\sum_{q=0}^{2s}\binom{2s}{q}(\xi_+\xi_-)^q  K_q(k)K_q(\ell) = \frac{\displaystyle (1+\xi_+\xi_-)^{2s}}{\displaystyle (\xi_+\xi_-)^{k}\binom{2s}{k}} \ \delta_{k,\ell},\label{eq:ort1}\\
&\sum_{q=0}^{2s}\binom{2s}{q}(\xi_+\xi_-)^q \,q\, K_q^2 = \frac{\displaystyle (1+\xi_+\xi_-)^{2s-1}}{\displaystyle (\xi_+\xi_-)^{k}\binom{2s}{k}}(k+(2s-k)\xi_+\xi_-),\label{eq:ort2}\\
&\sum_{q=0}^{2s}\binom{2s}{q}(\xi_+\xi_-)^q \,q\,   K_qK_q(k-1) = -\frac{\displaystyle (1+\xi_+\xi_-)^{2s-1}}{\displaystyle (\xi_+\xi_-)^{k-1}\binom{2s}{k}}(2s-k+1),\label{eq:ort3}\\
&\sum_{q=0}^{2s}\binom{2s}{q}(\xi_+\xi_-)^q \,q^2\, K_q^2 = \frac{\displaystyle (1+\xi_+\xi_-)^{2s-2}}{\displaystyle(\xi_+\xi_-)^{k}\binom{2s}{k}}
\left( (\xi_+\xi_-)^2(k-2s)^2 +2\xi_+\xi_-(4sk-2k^2+s)+k^2\right).\label{eq:ort4}
\end{align}
\end{lemme}
\textbf{Proof}.
Relation (\ref{eq:ort1}) is directly obtained from \cite{Koekoek1994}.
Let us now differentiate the orthogonality relation (\ref{eq:ort1}) (for $\ell=k$) with respect to $\xi_+$. By using (\ref{eq:A7}), one gets
\begin{multline}
\sum_{q=0}^{2s}\binom{2s}{q}q\xi_+^{q-1}\xi_-^{q}K_q^2 - \frac{\displaystyle 2k}{\displaystyle \xi_+(1+\xi_+\xi_-)} \sum_{q=0}^{2s}\binom{2s}{q}(\xi_+\xi_-)^qK_q(K_q-K_q(k-1))\\
= \frac{\displaystyle (1+\xi_+\xi_-)^{2s-1}\left[2s\xi_+\xi_- - k(1+\xi_+\xi_-)\right]}{\displaystyle \xi_+(\xi_+\xi_-)^{k}\binom{2s}{k}}.
\end{multline}
Therefore, by again using  (\ref{eq:ort1}), we get  (\ref{eq:ort2}).
Similarly, by differentiating the orthogonality relation (\ref{eq:ort1}) with respect to $\xi_+$ (for $\ell=k-1$), one gets \eqref{eq:ort3}.
Finally, we differentiate relation (\ref{eq:ort2}) to get relation (\ref{eq:ort4}).
$\left.\right.$\hfill$\square$

Let us recall that the Krawtchouk polynomials are self-dual \textit{i.e.} they satisfy
\begin{equation}
 K_j(k)=K_k(j).
\end{equation}
Therefore, from Lemma \ref{lem:ort}, one gets the other orthogonality properties
\begin{align}
&\sum_{k=0}^{2s}\binom{2s}{k}(\xi_+\xi_-)^k  K_j(k)K_\ell(k) = \frac{\displaystyle (1+\xi_+\xi_-)^{2s}}{\displaystyle (\xi_+\xi_-)^{j}\binom{2s}{j} }\  \ \delta_{j,\ell},\label{eq:ortd1}\\
&\sum_{k=0}^{2s}\binom{2s}{k}(\xi_+\xi_-)^k \,k\, K_j^2(k) = \frac{\displaystyle  (1+\xi_+\xi_-)^{2s-1}}{\displaystyle  (\xi_+\xi_-)^{j}\binom{2s}{j} }(j+(2s-j)\xi_+\xi_-),\label{eq:ortd2}\\
&\sum_{k=0}^{2s}\binom{2s}{k}(\xi_+\xi_-)^k \,k\,   K_j(k)K_{j-1}(k) = -\frac{\displaystyle  (1+\xi_+\xi_-)^{2s-1}}{\displaystyle  (\xi_+\xi_-)^{j-1}\binom{2s}{j}}(2s-j+1).\label{eq:ortd3}
\end{align}
Let us also remark that 
\begin{equation}
 \sum_{k=0}^{2s}\binom{2s}{k}(\xi_+\xi_-)^k \,k\,   K_j(k)K_{\ell}(k) = 0 \qquad\text{if}\quad \ell\neq k,k\pm1\ .\label{eq:ortd4}
\end{equation}
In our notation, the difference equation satisfied by the Krawtchouk polynomials (see (1.10.5) of \cite{Koekoek1994}) read
\begin{equation}\label{diff}
 -p(2s-k)K_j(k+1)+(k-j+2p(s-k))K_j-k(1-p)K_j(k-1)=0\;.
\end{equation}

\begin{lemme} The following relation between the Krawtchouk polynomials holds
\begin{equation}\label{eq:d4}
\frac{\displaystyle  1}{\displaystyle  1+\xi_+\xi_-}\left[ 2(s-j)K_j+\xi_+\xi_-(2s-j)K_{j+1}-\frac{\displaystyle  j}{\displaystyle  \xi_+\xi_-}K_{j-1} \right] =(2s-k)K_j(k+1).
\end{equation}
\end{lemme}
\textbf{Proof}.
By using the following properties of the hypergeometric functions  \cite{Wolfram}
\begin{eqnarray}
(b-c){}_{2}F_{1}\left(a, b-1; c; x\right)+(c-a-b){}_{2}F_{1}\left(a, b; c; x\right) = a(x-1){}_{2}F_{1}\left(a+1, b; c; x\right).
\end{eqnarray}
one deduces that
\begin{equation}
 K_j(k+1)=\frac{\displaystyle  1}{\displaystyle  2s-k}\left[ (2s-j-k)K_j-\frac{\displaystyle j}{\displaystyle  \xi_+\xi_-}K_{j-1} \right].
\end{equation}
Then, by replacing $K_j(k+1)$ in (\ref{eq:d4}) in the above formula, we see that (\ref{eq:d4}) is equivalent to the recurrence relation 
of the Krawtchouk polynomial (see (1.10.3) of \cite{Koekoek1994}) which concludes the proof.
$\left.\right.$\hfill$\square$

\end{document}